\documentclass[journal=jacsat,manuscript=article]{achemso}
\usepackage[T1]{fontenc}
\usepackage[utf8]{inputenc}
\usepackage{verbatim}
\usepackage{textcomp}
\usepackage{amssymb}
\SectionNumbersOn
\usepackage[version=3]{mhchem}
\usepackage{graphicx}
\usepackage{babel}

\global\long\def\ket#1{\left|#1\right\rangle }%
\global\long\def\bra#1{\left\langle #1\right|}%

\title{Atomistic theory of hot carrier relaxation in large plasmonic nanoparticles}
\author{S. M. João}
\affiliation{Department of Materials, Imperial College London, South Kensington Campus, London SW7 2AZ, United Kingdom}
\author{H. Jin}
\affiliation{Department of Materials, Imperial College London, South Kensington Campus, London SW7 2AZ, United Kingdom}
\author{J. Lischner}
\affiliation{Department of Materials, Imperial College London, South Kensington Campus, London SW7 2AZ, United Kingdom}
\email{j.lischner@imperial.ac.uk}

\begin{document}

\begin{abstract}
Recently, there has been significant interest in harnessing hot carriers generated from the decay of localized surface plasmons in metallic nanoparticles for applications in photocatalysis, photovoltaics and sensing. In this work, we present an atomistic approach to predict the population of hot carriers under continuous wave illumination in large nanoparticles. For this, we solve the equation of motion of the density matrix taking into account both excitation of hot carriers as well as subsequent relaxation effects. We present results for spherical Au and Ag nanoparticles with up to $250,000$ atoms. We find that the population of highly energetic carriers depends both on the material and the nanoparticle size. We also study the increase in the electronic temperature upon illumination and find that Ag nanoparticles exhibit a much larger temperature increase than Au nanoparticles. Finally, we investigate the effect of using different models for the relaxation matrix but find that qualitative features of the hot-carrier population are robust. 
\end{abstract}
\maketitle

\section{Introduction}


Metallic nanoparticles (NPs) feature strong and highly tunable light-matter interactions which makes them attractive for applications in photovoltaics and photocatalysis \citep{brongersmaPlasmoninducedHotCarrier2015,mukherjeeHotElectronsImpossible2013,bernardiInitioStudyHot2014,dalfornoMaterialSizeEnvironment2018,cortes_challenges_2020}. In particular, such NPs exhibit collective charge oscillations known as localized surface plasmons (LSPs) \citep{maierPlasmonicsFundamentalsApplications2007} which give rise to a pronounced peak in the optical absorption spectrum at the localized surface plasmon resonance (LSPR) frequency. LSPs are strongly damped oscillations that can decay via the Landau damping mechanism into energetic electron-hole pairs, known as hot carriers, which can then be harnessed for different applications. 

After their generation \citep{romancastellanosSinglePlasmonHot2019}, hot carriers lose their energy as a result of interactions with other electrons, lattice vibrations \citep{bernardiInitioStudyHot2014,giannini_controlling_2010,fitzgerald_quantum_2016} or defects. In addition, they can also interact with molecular species adsorbed to the surface of the NP which in turn can induce adsorbate vibrations, desorption events or chemical transformations. A detailed understanding of how these various processes influence and depend on the population of hot carriers is important for optimizing devices. 

To gain microscopic insights into hot-carrier relaxation processes, various theoretical approaches have been developed. For example, Dubi and Sivan \citep{dubiHotElectronsMetallic2019} solved the Boltzmann equation to calculate the steady-state population of hot carriers taking both electron-electron and electron-phonon scattering effects into account. A fully quantum-mechanical approach based on the equation of motion for the electronic density matrix was developed Govorov and coworkers \citep{govorovKineticDensityFunctional2015,govorovTheoryPhotoinjectionHot2013} who employed a relaxation time approximation. Another way to include relaxation effects is by solving a master equation for the electronic occupations of the NP states. Electron-phonon, electron-electron and electron-photon scattering mechanisms can be included, and their relative importance estimated \citep{saavedraHotElectronDynamicsThermalization2016,liu_relaxation_2018}.


While the aforementioned modelling efforts allowed important insights into the relaxation dynamics of hot carriers, it is important to note that they all employed highly simplified models of the NP electronic structure: they often use simple spherical well models or assume that the nanoparticle's density of states is the same as that of a bulk free electron gas. As a consequence, such models cannot capture hot carriers generated in  $d$ bands \citep{romancastellanosGenerationPlasmonicHot2020}, describe the dependence of the electronic structure on the exposed facets of the NP \citep{rossiHotCarrierGenerationPlasmonic2020a}, or describe plasmonic heterostructures atomistically \citep{jin_theory_2023}, all of which can play a critical role \citep{al-zubeidi_d-band_2023,bykov_dynamics_2021, brown_nonradiative_2016,kiani_interfacial_2023,ramakrishnan_photoinduced_2021}.


In this paper, we start from the density matrix formalism developed in \citep{govorovKineticDensityFunctional2015} and combine it with an atomistic model of the NP's electronic structure based on the tight-binding approach. To enable the simulation of large NPs, we cast the expression of the hot-carrier population in a basis-independent way and use the Kernel Polynomial Method (KPM) \citep{weisseKernelPolynomialMethod2006,silverDensitiesStatesMegaDimensional1994} to evaluate it. In recent years, KPM and other spectral methods have been established as promising tools to study quantum-mechanical phenomena in mesoscopic systems. For example, they have been used to extract spectral properties \citep{joaoHighresolutionRealspaceEvaluation2022,piresBreakdownUniversalityThreedimensional2021,PhysRevLett.115.106601} and transport properties \citep{joaoKITEHighperformanceAccurate2020,weisse2004chebyshev,santospiresAnomalousTransportSignatures2022,joaoBasisindependentSpectralMethods2019,covaciEfficientNumericalApproach2010,PhysRevMaterials.2.034004,canonicoShubnikovHaasOscillations2018} in multi-billion atom lattices. Recently, they have also been leveraged to obtain the rate of hot-carrier generation in million-atom plasmonic nanoparticles \citep{jinPlasmonInducedHotCarriers2022}. Here, we use this approach to study the hot-carrier population in large spherical Au and Ag nanoparticles.

The paper is organized as follows. In Section \ref{sec:Theory-and-computational} we explain how the optical properties of the NP, the electronic Hamiltonian and the density matrix are constructed before using these quantities to determine the non-equilibrium steady-state hot-carrier population. In Section \ref{sec:Results-and-Discussion}, we discuss the sensitivity of the hot-carrier population on the values of the relaxation times and then study the dependence of the hot-carrier population on the diameter of spherical Au and Ag NPs. This section ends with a discussion of the temperature increase in illuminated NPs.

\section{Theory and computational details\label{sec:Theory-and-computational}}

In this section, we describe our theoretical and computational methodology. For further details, see Ref. \citep{jinPlasmonInducedHotCarriers2022}. We start with the optical properties of the NP, discuss the electronic Hamiltonian and density matrix and finally show how the steady-state population can be obtained in a basis-independent way.

\subsection{Optical properties}

The interaction of the NP with light is modeled by considering the effect of a monochromatic external electric field of angular frequency $\omega$ given by $\mathbf{E}_{\text{ext}}\left(\mathbf{r},t\right)=E_{0}\cos\left(\omega t-kz\right)\hat{\mathbf{z}}$, where $k\hat{\mathbf{z}}$ denotes the wave vector  and $E_0$ is the amplitude. In this paper we are interested in NPs whose diameters range from $\sim2$ nm to $\sim20$ nm, which are much smaller than the typical wavelength of the external electric fields. In this case $\mathbf{E}_{\text{ext}}$ can be considered uniform inside the NP. This is the quasistatic approximation (QSA). To calculate the total electric field experienced by the electrons inside the NP, it is then sufficient to solve Laplace's equation $\nabla\cdot (\varepsilon(\mathbf{r},\omega)\nabla\phi(\mathbf{r},\omega))=0$ for the total electric potential $\phi$, with boundary condition at infinity $\phi\left(\mathbf{r},t\right)=-E_{0}z\cos\left(\omega t\right)$ and appropriate dielectric functions $\varepsilon(\omega)$ in each region. Here we are only considering spherical nanoparticles for which the analytical solution of Laplace's equation is given by 

\[
\phi\left(\mathbf{r},t\right)=\frac{3\varepsilon_{m}}{2\varepsilon_{m}+\varepsilon\left(\omega\right)}\left(-E_{0}z\right)\cos\left(\omega t\right)=\phi\left(\mathbf{r}\right)\cos\left(\omega t\right),
\]
where $\varepsilon_{m}$ is the dielectric constant of the environment and the measured dielectric function of the bulk material (either Ag or Au) were taken from Ref. \citep{haynesCRCHandbookChemistry2014}. Here, we consider NPs in vacuum and set $\varepsilon_{m}=\varepsilon_{0}$, the vacuum permittivity. The factor
\[
\alpha(\omega) = \left|\frac{3\varepsilon_{m}}{2\varepsilon_{m}+\varepsilon\left(\omega\right)}\right|\label{intensity prefactor}
\]
reflects the enhancement of the electric field inside the NP. The potential energy $\Phi\left(\mathbf{r},t\right)$ of an electron at position $\mathbf{r}$ and time $t$ is given by $\Phi\left(\mathbf{r},t\right)=-e\phi\left(\mathbf{r},t\right)$, where $-e<0$ is the electron's charge. This optical perturbation excites the electron-hole pairs which then interact with phonons and other electrons and relax to a non-equilibrium  population under continuous wave illumination. 

\subsection{Electronic Hamiltonian}

The dynamics of electrons is described by a tight-binding model $H=H_{0}+\Phi\left(t\right)$, where
\[
H_{0}=\sum_{ij\alpha\beta}t_{ij}^{\alpha\beta}c_{i,\alpha}^{\dagger}c_{j,\beta}
\]
represents the Hamiltonian in the absence of illumination and
\[
\Phi\left(t\right)=\sum_{i\alpha}\Phi\left(\mathbf{R}_{i},t\right)c_{i,\alpha}^{\dagger}c_{i,\alpha}
\]
describes the optical perturbation. Here, $c_{i,\alpha}^{\dagger}$ ($c_{i,\alpha}$) denotes the creation (annihilation) operator of an electron in orbital $\alpha$ of an atom at position $\mathbf{R}_{i}$, $t_{ij}^{\alpha\beta}$ is the hopping and onsite matrix and the double sum over $i$ and $j$ is typically constrained to a small number of neighbours, making $t_{ij}^{\alpha\beta}$ a sparse matrix. The tight-binding models used in this paper are obtained through an orthogonal two-center Slater-Koster parameterization \citep{papaconstantopoulosHandbookBandStructure2015}. For Au, we use a parameterization consisting of $5d$, $6s$ and $6p$ atomic orbitals and for Ag, we use $4d$, $5s$ and $5p$ orbitals.

\subsection{Density matrix}

To determine the non-equilibrium population of electrons inside the NP, we solve the equation of motion for the electronic density matrix $\rho\left(t\right)$ given by~\citep{govorovKineticDensityFunctional2015}

\[
-i\hbar\frac{\partial\rho_{mn}}{\partial t}=\bra m\left[\rho,H_{0}+\Phi\left(t\right)\right]\ket n+i\Gamma_{mn}\delta\rho_{mn},
\]
where $|n\rangle$ denotes an eigenstate of $H_0$ with corresponding eigenenergy $\varepsilon_n$ and $\rho_{mn}=\langle m | \rho | n \rangle$. Relaxation effects are described phenomenologically through the matrix $\Gamma_{mn}$. Also, $\delta\rho_{mn}=\rho_{mn}-\rho_{mn}^{0}$, where $\rho_{mn}^{0}=\delta_{nm}f\left(\varepsilon_{n}\right)$ is the density matrix before the perturbation has been turned on, given by the Fermi-Dirac distribution $f\left(\varepsilon\right)=\left(1+e^{\beta\left(\varepsilon-\mu\right)}\right)^{-1}$, with chemical potential $\mu$ and $\beta=1/(k_B T)$ (with $k_B$ and $T$ denoting the Boltzmann constant and the temperature, respectively). For all our calculations, we use room temperature ($T=298K$).

The equation of motion for the density matrix can be solved using perturbation theory assuming a weak monochromatic perturbation $\Phi\left(t\right)=\varphi_{\omega}e^{i\omega t}+\varphi_{\omega}^{\dagger}e^{-i\omega t}$ \citep{govorovKineticDensityFunctional2015}. The result can be used to calculate the excess density of electrons with energy $E$, $\delta N\left(E\right)=\sum_{n}\delta \rho_{nn}\delta\left(E-\varepsilon_{n}\right)$, relative to the equilibrium value. This yields

\begin{eqnarray}
 &  & \delta N\left(E\right)=-\sum_{n}\delta\left(E-\varepsilon_{n}\right)\frac{4}{\Gamma_{nn}}\sum_{m}\left(f_{n}-f_{m}\right)\label{eq: distribution eigen}\\
 &  & \quad\quad\times\left[\frac{\Gamma_{nm}\left|\varphi_{\omega,mn}\right|^{2}}{\left(\hbar\omega-\varepsilon_{m}+\varepsilon_{n}\right)^{2}+\Gamma_{nm}^{2}}+\frac{\Gamma_{nm}\left|\varphi_{\omega,mn}\right|^{2}}{\left(\hbar\omega+\varepsilon_{m}-\varepsilon_{n}\right)^{2}+\Gamma_{nm}^{2}}\right].\nonumber 
\end{eqnarray}

Evaluating this expression requires knowledge of the eigenfunctions and eigenvalues of the Hamiltonian $H_{0}$. Calculating these quantities using a diagonalization procedure becomes numerically challenging for large NPs of relevance to nanoplasmonic experiments.

\subsection{Basis-independent population \label{subsec:Basis-independent-distribution}}

To study large NPs, it is useful to express Eq. \ref{eq: distribution eigen} in a basis-independent form. To achieve this, we first assume that $\Gamma$ can be expressed as a function of the eigenenergies, i.e. $\Gamma_{nm}=\Gamma\left(\varepsilon_{n},\varepsilon_{m}\right)$. Then, following \citep{jinPlasmonInducedHotCarriers2022} and introducing a double integral over $\varepsilon$ and $\varepsilon^{\prime}$ weighted by $\delta\left(\varepsilon-\varepsilon_{n}\right)\delta\left(\varepsilon^{\prime}-\varepsilon_{m}\right)$, we find
\begin{equation}
\delta N\left(E\right)=\int_{-\infty}^{\infty}d\varepsilon\int_{-\infty}^{\infty}d\varepsilon^{\prime}\delta\left(E-\varepsilon\right)\Omega\left(\varepsilon,\varepsilon^{\prime}\right)\Lambda\left(\varepsilon,\varepsilon^{\prime}\right),\label{eq: distribution real space}
\end{equation}

with
\[
\Lambda\left(\varepsilon,\varepsilon^{\prime}\right)=-4\frac{f\left(\varepsilon\right)-f\left(\varepsilon^{\prime}\right)}{\Gamma\left(\varepsilon,\varepsilon\right)}\left[\frac{\Gamma\left(\varepsilon,\varepsilon^{\prime}\right)}{\left(\hbar\omega-\varepsilon^{\prime}+\varepsilon\right)^{2}+\Gamma^{2}\left(\varepsilon,\varepsilon^{\prime}\right)}+\frac{\Gamma\left(\varepsilon,\varepsilon^{\prime}\right)}{\left(\hbar\omega+\varepsilon^{\prime}-\varepsilon\right)^{2}+\Gamma^{2}\left(\varepsilon,\varepsilon^{\prime}\right)}\right]
\]
and
\[
\Omega\left(\varepsilon,\varepsilon^{\prime}\right)=\text{Tr}\left[\delta\left(\varepsilon-H_{0}\right)\varphi_{\omega}^{\dagger}\delta\left(\varepsilon^{\prime}-H_{0}\right)\varphi_{\omega}\right].
\]

Note that $\Lambda\left(\varepsilon,\varepsilon^{\prime}\right)$ contains all information about relaxation processes through $\Gamma$. The Lorentzian form of the term inside the square brackets ensures energy conservation in optical transitions from $\varepsilon$ to $\varepsilon^{\prime}$ with a finite linewidth $\Gamma\left(\varepsilon,\varepsilon^{\prime}\right)$. The difference of Fermi functions ensures that only transitions from occupied states to unoccupied state are allowed. $\Omega\left(\varepsilon,\varepsilon^{\prime}\right)$ is the energy-resolved optical transition matrix. Importantly, Eq. \ref{eq: distribution real space} factorizes the information about statistics and relaxation processes in $\Lambda\left(\varepsilon,\varepsilon^{\prime}\right)$ and the information about the single-particle electron dynamics in $\Omega\left(\varepsilon,\varepsilon^{\prime}\right)$. Typically, $\Omega\left(\varepsilon,\varepsilon^{\prime}\right)$ is the most difficult object to calculate, but once it is obtained, any relaxation matrix can be studied without additional computational cost as a consequence of the factorization. $\Omega\left(\varepsilon,\varepsilon^{\prime}\right)$ is calculated efficiently using the Kernel Polynomial Method \citep{weisseKernelPolynomialMethod2006} in the same way as in Ref \citep{jinPlasmonInducedHotCarriers2022}.

\section{Results and Discussion \label{sec:Results-and-Discussion}}

In this section, we calculate the steady-state electron population in spherical Au and Ag NPs of different sizes. We also explore different simple models for the relaxation matrix, but find that the gross features of the electron population do not depend on the model for the relaxation matrix.

\subsection{Relaxation matrix model}

The relaxation matrix $\Gamma\left(\varepsilon,\varepsilon^{\prime}\right)$ is one of the central objects in Eq. \ref{eq: distribution real space}, and its functional form depends on the details of electron-electron and electron-phonon scattering mechanisms. While our approach allows a considerable amount of flexibility in choosing $\Gamma$, here we explore a few simple models to understand their effect on the steady-state electron population of spherical Au and Ag NPs. Following \citep{govorovKineticDensityFunctional2015}, we parameterize the matrix by the energy relaxation time $\tau_{\varepsilon}$ ($\tau_\varepsilon=1.3\text{ meV}$ for both Au and Ag) and the momentum relaxation time $\tau_{p}$ ($\tau_p=78 \text{ meV}$ for Au and $\tau_p=20 \text{ meV}$ for Ag). The momentum relaxation time is obtained from the Drude fit to the optical constants of Au and Ag obtained experimentally \citep{johnson_optical_1972}. The energy relaxation time estimates the electronic thermalization time and has considerable contributions coming from both electron-electron and electron-phonon scattering. Pump-probe experiments provide information about both these processes and place the corresponding times in the order of magnitude of $0.1-1\text{ ps}$ \citep{linkSpectralPropertiesRelaxation1999,voisin_size-dependent_2000,arbouet_electron-phonon_2003}, depending on the experimental apparatus, NP sizes and frequency at which the measurement was made (among other factors). For simplicity, we consider their collective effect to yield $\tau_{\varepsilon}\approx0.5\text{ ps}$. In \citep{govorovKineticDensityFunctional2015}, $\Gamma_{nn}=\tau_\varepsilon$ and $\Gamma_{nm}=\tau_p$ if $m \neq n$. To mimic this form for the relaxation matrix, we assign $\tau_{\varepsilon}$ to transitions which have very similar energies, that is $\left|\varepsilon_{n}-\varepsilon_{m}\right|<20\text{ meV}$, and assign $\tau_{p}$ otherwise. This relaxation matrix is referred to as $\Gamma_1$ and given by

\[
\Gamma_1\left(\varepsilon,\varepsilon^{\prime}\right)=\begin{cases}
\tau_{\varepsilon} & \text{if } \left|\varepsilon-\varepsilon^{\prime}\right|<20\text{ meV}\\
\tau_{p} & \text{if } \left|\varepsilon-\varepsilon^{\prime}\right|\geq20\text{ meV}
\end{cases}.
\]

To test the sensitivity of our results to the numerical values of the relaxation times, we also present results for a relaxation matrix in which the value of $\tau_p$ is doubled (denoted by $\Gamma_2$) and a relaxation matrix in which the value of $\tau_\varepsilon$ is doubled (denoted by $\Gamma_3$).

\begin{figure}
\includegraphics[scale=0.6]{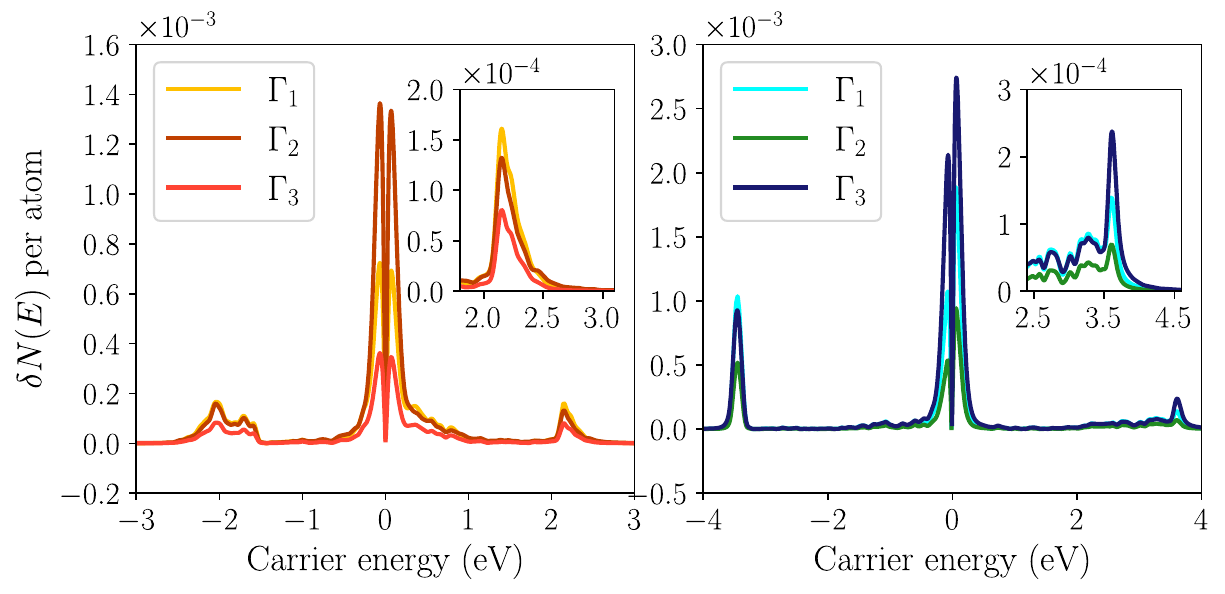}
\caption{Steady-state hot-carrier population in spherical Au (left) and Ag (right) NPs with a diameter of $20$ nm for three different relaxation matrices ($\Gamma_{1}$, $\Gamma_{2}$, $\Gamma_{3}$) at their corresponding LSPR ($2.4$ eV for Au, $3.5$ eV for Ag) at room temperature ($T=298K$). The Fermi energy is set to zero, and the hole contribution to the hot-carrier population has been flipped in sign for clarity. The inset shows the high energy peak of the hot-electron population.}
\label{Fig: different matrices}
\end{figure}

Figure \ref{Fig: different matrices} compares the carrier population in Au and Ag NPs at their LSPR frequencies for the three relaxation matrices under an external electric field $\left|\mathbf{E}_{\text{ext}}\right|=8.7\times10^{-4}$ V/nm, corresponding to an illumination power of $1 \text{mW}/\mu\text{m}^2$. Carriers with positive (negative) energy correspond to electrons (holes). The sign of the hole contribution has been inverted for clarity. Transitions from the $d$ band to the $sp$ band generate hot holes and cold electrons in both metals, giving rise to a hole peak at the position of the $d$ bands ($-2$ eV for Au, $-3.5$ eV for Ag) and a corresponding electron peak close to the Fermi level. Transitions in which both the initial and the final states derive from the $sp$ band of the bulk material (also known as intraband transitions) are responsible for the electron peak at $2.2$ eV in Au and the smaller electron peak 3.5 eV in Ag. The large number of electrons and holes near the Fermi level is due to relaxation effects which drive the hot electrons and holes back towards thermal equilibrium. See section \ref{subsec:Size-effect} for more details on interpreting these curves.

Comparing the results of the three different relaxation matrix we observe that the energy relaxation time mostly determines the overall magnitude of $\delta N$: the populations obtained with $\Gamma_1$ and $\Gamma_3$ (which have the same momentum relaxation time) have the same shape, but differ in their magnitude. This can be understood from Eq. \ref{eq: distribution eigen}, since $1/\Gamma_{nn}=1/\tau_\varepsilon$ enters as an overall prefactor. In contrast, the shape of $\delta N$ is affected by the momentum relaxation time, especially around the Fermi level. However, we find that the qualitative features of the electron population are the same for all relaxation models and we therefore present results only for $\Gamma_1$ in the remainder of this paper.

\subsection{Size dependence \label{subsec:Size-effect}}

\begin{figure}
\includegraphics[scale=0.6]{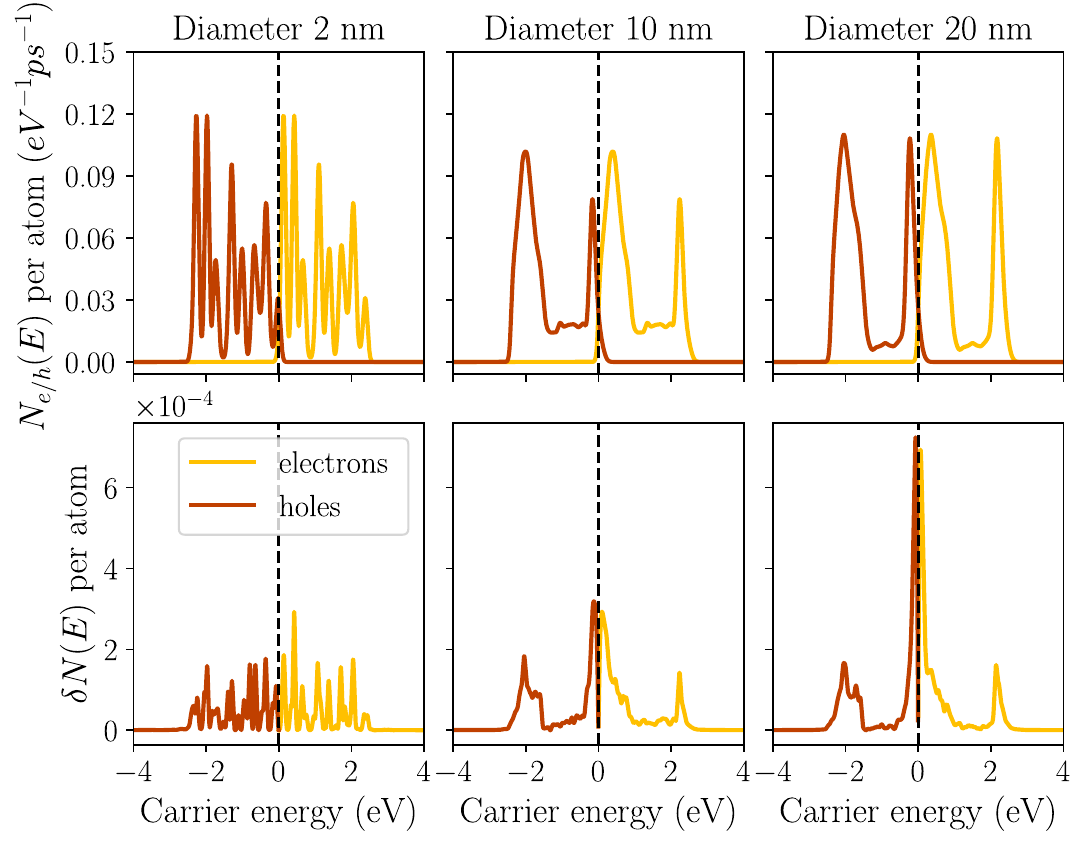}
\caption{Hot-carrier generation rate (top) and population (bottom) of spherical Au NPs at the LSPR frequency ($\hbar\omega=2.4$ eV) for three different diameters at room temperature ($T=298K$).}
\label{Fig: Au HCG + dist}
\end{figure}

In this section, we explore the effect of NP size on the steady-state population. Fig. \ref{Fig: Au HCG + dist} shows the rate of carrier generation in Au NPs, obtained using the method from \citep{jinPlasmonInducedHotCarriers2022}, as well as the electron and hole population inside the NP after relaxation, at the LSPR frequency $\hbar\omega=2.4$ eV. 

\begin{figure}
\includegraphics[scale=0.6]{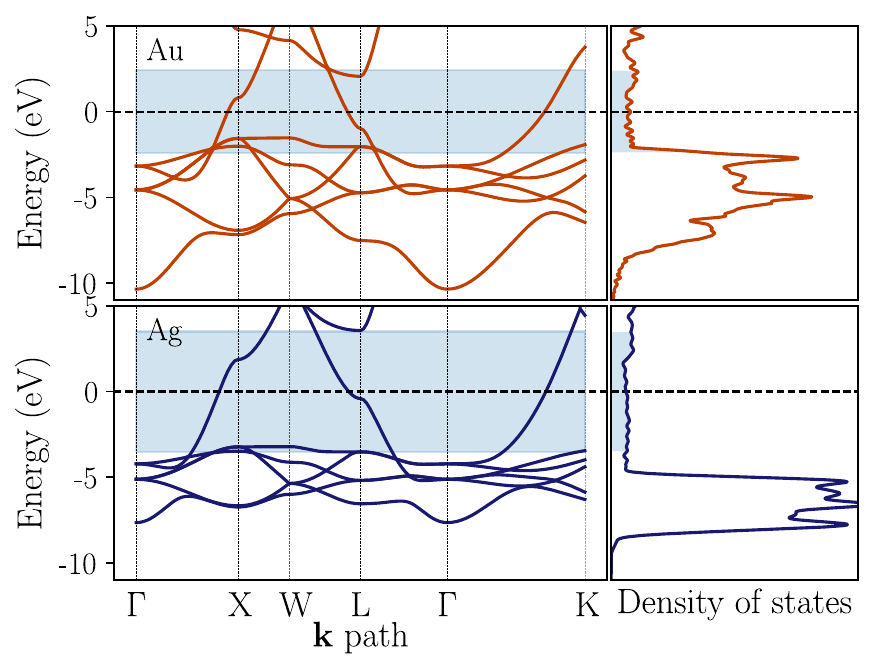}
\caption{Band structure and density of states of Au and Ag, using the tight-binding model from Ref. \citep{papaconstantopoulosHandbookBandStructure2015}. The state that can participate in optical transitions at the LSPR frequency lie within the blue rectangle, and the Fermi energy is set to zero.}
\label{Fig: band structures}
\end{figure}

The hot-carrier generation rates have contributions both from momentum-conserving (direct) and non-conserving (indirect) optical transitions. Direct transitions can be understood in terms of gold's band structure (see top panel of Fig \ref{Fig: band structures}).  There is a large number of transitions arising from the $d$ bands at $E=-2$ eV, giving rise to a large number of hot holes at $-2$ eV and cold electrons at $0.4$ eV, close to the Fermi level. A similar situation occurs at $E=2.2$ eV, at the onset of an $sp$ band. These transitions generate many hot electrons but relatively cold holes at $-0.2$ eV. Indirect transitions do not conserve momentum and the typical scale of momentum transferred is inversely proportional to the NP diameter \citep{govorovTheoryPhotoinjectionHot2013}. For this reason, the relative importance of indirect transitions increases as the NP size decreases. For very small nanoparticles (see results for 2 nm diameter), the generation rate exhibits a molecule-like behaviour with many discrete peaks arising from quantum confinement effects.

The steady-state hot-carrier population exhibits a similar peak structure as the generation rate, but the relative height of the peaks is different compared to the generation rate: after the carriers are generated, they relax via electron-electron and electron-phonon interactions, losing energy and approaching the Fermi level. For example, electrons are constantly generated at $E=2.2$ eV contributing to the $E=2.2$ eV peak in the electron population (see bottom panel of Fig. \ref{Fig: Au HCG + dist}). However, the electrons constantly relax back to the Fermi energy and contribute to the peak near $E=0$ eV. A similar picture holds for the holes generated at $E=-2$ eV. 

To quantitatively study the number of high-energy electrons ($\delta N_\text{e}^\text{HE}$) and holes ($\delta N_\text{h}^\text{HE}$) in the steady-state population, we evaluate 
\begin{eqnarray}
\delta N^\text{HE}_{\text{h}} & = & \int_{-\infty}^{-E_\text{thr}} dE \delta N(E)\label{eq: integrated hot elecs}\\
\delta N^\text{HE}_{e} &=& \int^{\infty}_{E_\text{thr}} dE \delta N(E) ,\label{eq: integrated hot holes}
\end{eqnarray}
where $E_\text{thr}=1.5$~eV denotes a threshold energy. The value of the threshold energy is chosen such that the contribution from the high-energy peaks in $\delta N(E)$ is captured.  

\begin{figure}
\includegraphics[scale=0.6]{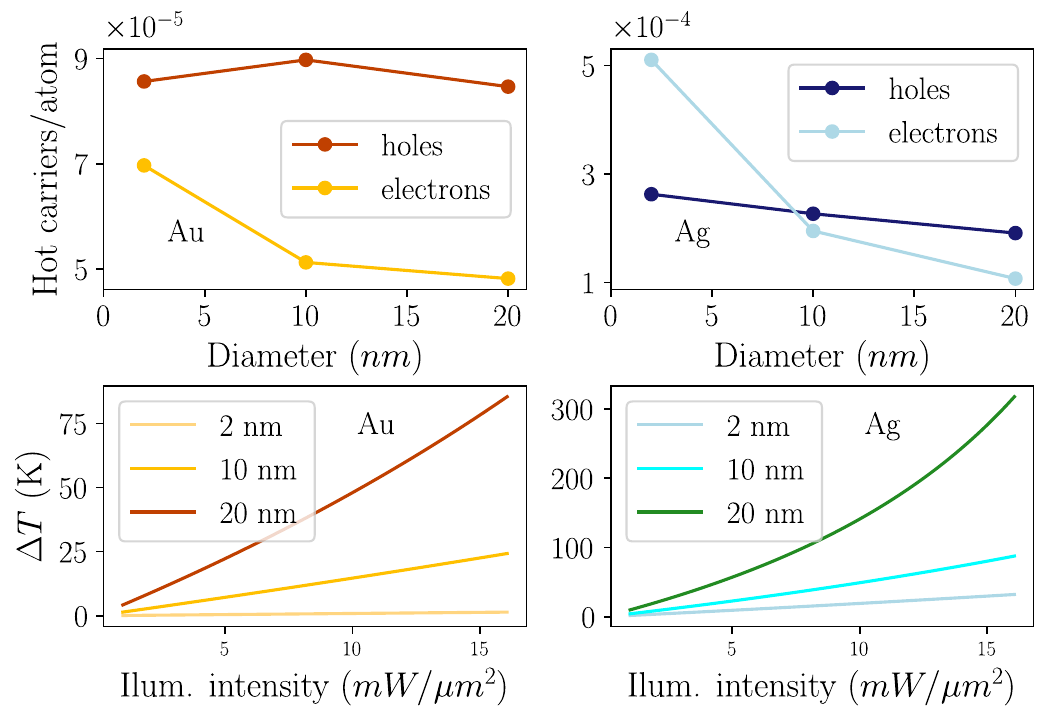}
\caption{Top panels: Number of high-energy electrons and holes in the steady-state population. In Au, high-energy carriers are defined to have an energy larger than $1.5$~eV relative to the Fermi level. In Ag, the corresponding energy threshold is $2.0$ eV. Bottom panels: Increase in electronic temperature as function of external illumination strength for different nanoparticle diameters.}
\label{Fig: Integrated hot carriers}
\end{figure}

Fig. \ref{Fig: Integrated hot carriers} (top left) shows that significantly more energetic holes are available in the steady-state population than energetic electrons, in particular for larger NPs. This asymmetry can be understood from the amount of available interband transitions. Fig. \ref{Fig: band structures} (top) shows that there is a much larger overlap of the LSPR zone with the $d$ bands at $-2$ eV than with the $sp$ band at $2$ eV. This asymmetry becomes more pronounced as the relative number of surface-enabled intraband transitions diminishes for larger NPs. 

\begin{figure}
\includegraphics[scale=0.6]{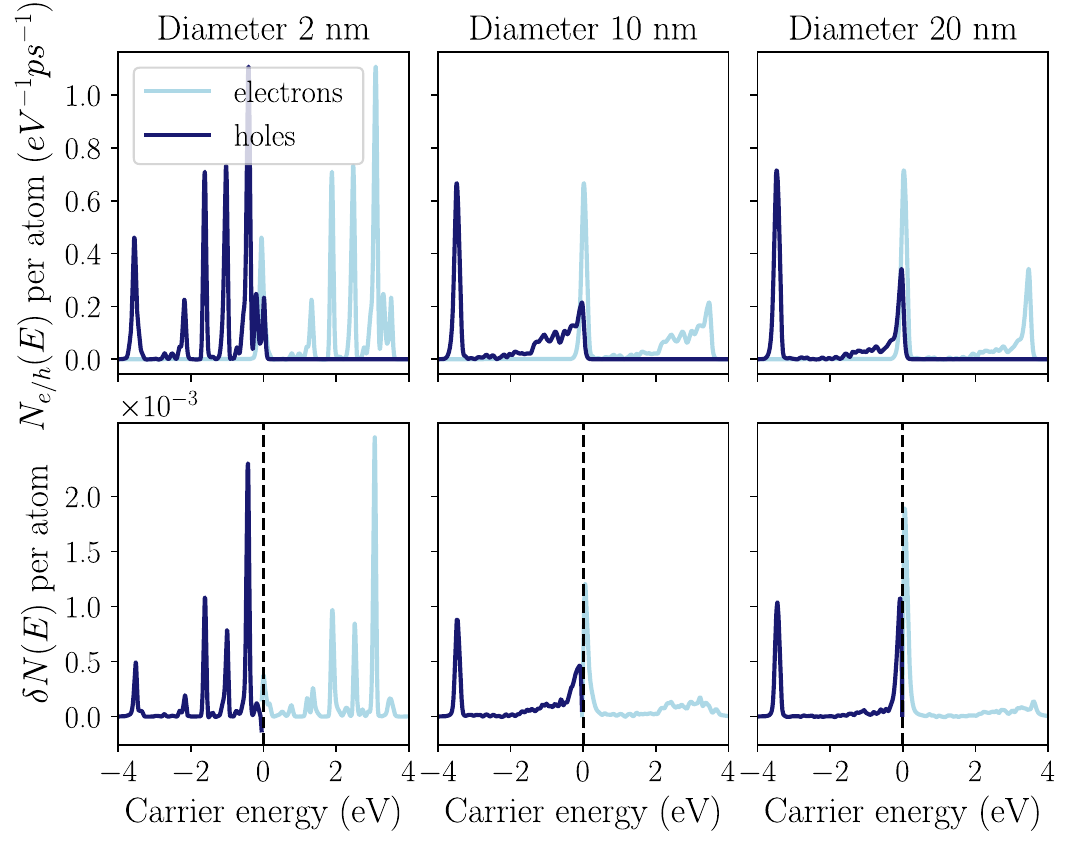}
\caption{Hot-carrier generation rates (top) and populations (bottom) for three different sizes of Ag NPs at the LSPR frequency ($\hbar\omega=3.5$ eV) at room temperature ($T=298K$).}
\label{Fig: Ag HCG + dist}
\end{figure}

Next, we repeat the analysis for Ag NPs. Ag has a qualitatively similar band structure to Au (see bottom panel of Fig. \ref{Fig: band structures}), but both the $d$ bands and the unoccupied $sp$ bands are further away from the Fermi level. The LSPR frequency is $\hbar\omega=3.5$ eV instead of $\hbar\omega=2.4$ eV. Compared to Au, the LSPR region has a smaller overlap with these bands in Ag (see Fig. \ref{Fig: band structures}), so we expect the high-energy peaks to be sharper in both the generation rate and the steady-state population due to the smaller number of available transitions. This is indeed what is observed in Fig. \ref{Fig: Ag HCG + dist}. The pronounced peak in the hot-hole generation at $-3$ eV persists in the hole population, but the sharp peak at $3.5$ eV in the hot-electron generation rate almost disappears after relaxation effects are taken into account. In its place, a relatively uniform and broad distribution of hot electrons appears.

Because of this sensitivity to the availability of compatible interband transitions in the hot-electron sector, there is a strong size dependency in the number of hot electrons $\delta N_\text{e}^\text{HE}$ to NP size, see Fig. \ref{Fig: Integrated hot carriers} (top right). In contrast, the number of high-energy holes is almost constant. Note that an energy threshold $E_\text{thr}=2.0$~eV was used in Eqs. \ref{eq: integrated hot elecs} and \ref{eq: integrated hot holes} for Ag NPs.

\subsection{Electronic temperature \label{subsec:temperature}}

 The nanoparticle's temperature is a key ingredient in explaining many chemical phenomena, and the electronic temperature can be estimated from energy dissipation with the heat equation \citep{movsesyanPlasmonicNanocrystalsComplex2022}, or by finding the solution to Boltzmann's equation in steady state \citep{dubiHotElectronsMetallic2019}. Our formalism also provides a way to compute the temperature of the electrons inside the nanoparticle via its connection to the Fermi function. In equilibrium, the electronic temperature $T$ can be determined from the slope of the Fermi-Dirac population at the Fermi level (which we set to zero here), i.e. $f'(0) = -1/(4k_B T)$ (with $k_B$ being the Boltzmann constant). Upon illumination, the electron population is no longer determined by a Fermi-Dirac distribution, but instead given by $ N(E)=\sum_n f(\epsilon_n) \delta(E-\epsilon_n) + \delta N(E)$, see Eq. ~\ref{eq: distribution eigen}. Note that near the Fermi level, we have $\sum_n f(\epsilon_n) \delta(E-\epsilon_n) \approx f(E) D(0)$ with $D(E)=\sum_n \delta(E-\epsilon_n)$ denoting the density of states which varies slowly near the Fermi level, see Fig. \ref{Fig: band structures}.

To extract an electronic temperature from $N(E)$, we assume that it behaves like an equilibrium population with an elevated temperature $T'$ near the Fermi level, i.e. $N(E) \approx f_{T'}(E)D(0)$ (where we now explicitly write the temperature as a subscript to the Fermi-Dirac function). Near the Fermi level, we thus have
$$D\left(0\right)f_{T^{\prime}}\left(E\right) =\delta N\left(E\right)+D\left(0\right)f_{T}\left(E\right).$$

The new temperature of the NP is then found to be
$$\frac{1}{T^{\prime}}=\frac{1}{T}-4k_{B}\frac{\delta N^{\prime}\left(0\right)}{D\left(0\right)}.$$

The increase in temperature as function of the illumination intensity is shown in Fig. \ref{Fig: Integrated hot carriers} (bottom). The temperature increase is mostly proportional to the field intensity. In larger NPs (most pronounced for Ag), a crossover to a parabolic dependence can be observed at larger illumination intensities. Larger NPs exhibit a larger increase in temperature compared to smaller NPs as there is a larger portion of electrons and holes close to the Fermi energy (see Figs. \ref{Fig: Au HCG + dist} and \ref{Fig: Ag HCG + dist}). Ag NPs show a much higher temperature increase than Au NPs for the same field strength. This is explained mainly by the observation that the field enhancement (given in Eq. \ref{intensity prefactor}) in Ag is much larger than in Au. Using the measured dielectric functions from \citep{haynesCRCHandbookChemistry2014}, we estimate $\alpha_{Au}=1.35$  for Au and $\alpha_{Ag}=5.02$ for Ag at the LSPR frequency.  The squared field enhancement enters $\delta N(E)$ explaining the large difference in temperature increase in Ag and Au NPs.

\section{Conclusion}

We have developed an atomistic approach to calculate the steady-state hot-carrier population in spherical Au and Ag NPs containing hundreds of thousands of atoms. This was achieved by expressing the population in a basis-independent way and using the Kernel Polynomial Method. Relaxation processes are included in this formalism through an energy-dependent relaxation matrix $\Gamma\left(\varepsilon,\varepsilon^{\prime}\right)$. We explored different simple models for the relaxation matrix, but found that the qualitative predictions do not depend on this choice. The hot-carrier populations of Ag and Au nanoparticles exhibit similar features, but the number of highly energetic electrons in Ag NPs found to be highly sensitive to the NP size. We explain this observation in terms of Ag's band structure which features an occupied band whose energy relative to the Fermi level is almost exactly the same as Ag LSP energy. Furthermore, the increase of the electron temperature in the NP is estimated and it is found that Ag NPs exhibit a much larger increase as a consequence of strong electric field enhancement inside the NP. The approach developed in this work can be used to study the hot-carrier population in NPs of complex shapes and compositions. This will be the focus of future work.

\begin{acknowledgement}

S.M.J. and J.L. acknowledge funding from the Royal Society through a Royal Society University Research Fellowship URF\textbackslash R\textbackslash 191004. J.L. acknowledges funding from the EPSRC programme grant EP/W017075/1. H.J. acknowledges financial support from his parents.
\end{acknowledgement}

\bibliography{all_library}

\providecommand{\latin}[1]{#1}
\makeatletter
\providecommand{\doi}
  {\begingroup\let\do\@makeother\dospecials
  \catcode`\{=1 \catcode`\}=2 \doi@aux}
\providecommand{\doi@aux}[1]{\endgroup\texttt{#1}}
\makeatother
\providecommand*\mcitethebibliography{\thebibliography}
\csname @ifundefined\endcsname{endmcitethebibliography}
  {\let\endmcitethebibliography\endthebibliography}{}
\begin{mcitethebibliography}{43}
\providecommand*\natexlab[1]{#1}
\providecommand*\mciteSetBstSublistMode[1]{}
\providecommand*\mciteSetBstMaxWidthForm[2]{}
\providecommand*\mciteBstWouldAddEndPuncttrue
  {\def\EndOfBibitem{\unskip.}}
\providecommand*\mciteBstWouldAddEndPunctfalse
  {\let\EndOfBibitem\relax}
\providecommand*\mciteSetBstMidEndSepPunct[3]{}
\providecommand*\mciteSetBstSublistLabelBeginEnd[3]{}
\providecommand*\EndOfBibitem{}
\mciteSetBstSublistMode{f}
\mciteSetBstMaxWidthForm{subitem}{(\alph{mcitesubitemcount})}
\mciteSetBstSublistLabelBeginEnd
  {\mcitemaxwidthsubitemform\space}
  {\relax}
  {\relax}

\bibitem[Brongersma \latin{et~al.}(2015)Brongersma, Halas, and
  Nordlander]{brongersmaPlasmoninducedHotCarrier2015}
Brongersma,~M.~L.; Halas,~N.~J.; Nordlander,~P. Plasmon-Induced Hot Carrier
  Science and Technology. \emph{Nature Nanotechnology} \textbf{2015},
  \emph{10}, 25--34\relax
\mciteBstWouldAddEndPuncttrue
\mciteSetBstMidEndSepPunct{\mcitedefaultmidpunct}
{\mcitedefaultendpunct}{\mcitedefaultseppunct}\relax
\EndOfBibitem
\bibitem[Mukherjee \latin{et~al.}(2013)Mukherjee, Libisch, Large, Neumann,
  Brown, Cheng, Lassiter, Carter, Nordlander, and
  Halas]{mukherjeeHotElectronsImpossible2013}
Mukherjee,~S.; Libisch,~F.; Large,~N.; Neumann,~O.; Brown,~L.~V.; Cheng,~J.;
  Lassiter,~J.~B.; Carter,~E.~A.; Nordlander,~P.; Halas,~N.~J. Hot {{Electrons
  Do}} the {{Impossible}}: {{Plasmon-Induced Dissociation}} of {{H2}} on
  {{Au}}. \emph{Nano Letters} \textbf{2013}, \emph{13}, 240--247\relax
\mciteBstWouldAddEndPuncttrue
\mciteSetBstMidEndSepPunct{\mcitedefaultmidpunct}
{\mcitedefaultendpunct}{\mcitedefaultseppunct}\relax
\EndOfBibitem
\bibitem[Bernardi \latin{et~al.}(2014)Bernardi, {Vigil-Fowler}, Lischner,
  Neaton, and Louie]{bernardiInitioStudyHot2014}
Bernardi,~M.; {Vigil-Fowler},~D.; Lischner,~J.; Neaton,~J.~B.; Louie,~S.~G. Ab
  {{Initio Study}} of {{Hot Carriers}} in the {{First Picosecond}} after
  {{Sunlight Absorption}} in {{Silicon}}. \emph{Physical Review Letters}
  \textbf{2014}, \emph{112}, 257402\relax
\mciteBstWouldAddEndPuncttrue
\mciteSetBstMidEndSepPunct{\mcitedefaultmidpunct}
{\mcitedefaultendpunct}{\mcitedefaultseppunct}\relax
\EndOfBibitem
\bibitem[Dal~Forno \latin{et~al.}(2018)Dal~Forno, Ranno, and
  Lischner]{dalfornoMaterialSizeEnvironment2018}
Dal~Forno,~S.; Ranno,~L.; Lischner,~J. Material, {{Size}}, and {{Environment
  Dependence}} of {{Plasmon-Induced Hot Carriers}} in {{Metallic
  Nanoparticles}}. \emph{The Journal of Physical Chemistry C} \textbf{2018},
  \emph{122}, 8517--8527\relax
\mciteBstWouldAddEndPuncttrue
\mciteSetBstMidEndSepPunct{\mcitedefaultmidpunct}
{\mcitedefaultendpunct}{\mcitedefaultseppunct}\relax
\EndOfBibitem
\bibitem[Cortés \latin{et~al.}(2020)Cortés, Besteiro, Alabastri, Baldi,
  Tagliabue, Demetriadou, and Narang]{cortes_challenges_2020}
Cortés,~E.; Besteiro,~L.~V.; Alabastri,~A.; Baldi,~A.; Tagliabue,~G.;
  Demetriadou,~A.; Narang,~P. Challenges in {Plasmonic} {Catalysis}. \emph{ACS
  Nano} \textbf{2020}, \emph{14}, 16202--16219\relax
\mciteBstWouldAddEndPuncttrue
\mciteSetBstMidEndSepPunct{\mcitedefaultmidpunct}
{\mcitedefaultendpunct}{\mcitedefaultseppunct}\relax
\EndOfBibitem
\bibitem[Maier(2007)]{maierPlasmonicsFundamentalsApplications2007}
Maier,~S. \emph{Plasmonics: {{Fundamentals}} and {{Applications}}}; {Springer},
  2007\relax
\mciteBstWouldAddEndPuncttrue
\mciteSetBstMidEndSepPunct{\mcitedefaultmidpunct}
{\mcitedefaultendpunct}{\mcitedefaultseppunct}\relax
\EndOfBibitem
\bibitem[Rom{\'a}n~Castellanos \latin{et~al.}(2019)Rom{\'a}n~Castellanos, Hess,
  and Lischner]{romancastellanosSinglePlasmonHot2019}
Rom{\'a}n~Castellanos,~L.; Hess,~O.; Lischner,~J. Single Plasmon Hot Carrier
  Generation in Metallic Nanoparticles. \emph{Communications Physics}
  \textbf{2019}, \emph{2}, 1--9\relax
\mciteBstWouldAddEndPuncttrue
\mciteSetBstMidEndSepPunct{\mcitedefaultmidpunct}
{\mcitedefaultendpunct}{\mcitedefaultseppunct}\relax
\EndOfBibitem
\bibitem[Giannini \latin{et~al.}(2010)Giannini, Fernández-Domínguez,
  Sonnefraud, Roschuk, Fernández-García, and
  Maier]{giannini_controlling_2010}
Giannini,~V.; Fernández-Domínguez,~A.~I.; Sonnefraud,~Y.; Roschuk,~T.;
  Fernández-García,~R.; Maier,~S.~A. Controlling {Light} {Localization} and
  {Light}–{Matter} {Interactions} with {Nanoplasmonics}. \emph{Small}
  \textbf{2010}, \emph{6}, 2498--2507\relax
\mciteBstWouldAddEndPuncttrue
\mciteSetBstMidEndSepPunct{\mcitedefaultmidpunct}
{\mcitedefaultendpunct}{\mcitedefaultseppunct}\relax
\EndOfBibitem
\bibitem[Fitzgerald \latin{et~al.}(2016)Fitzgerald, Narang, Craster, Maier, and
  Giannini]{fitzgerald_quantum_2016}
Fitzgerald,~J.~M.; Narang,~P.; Craster,~R.~V.; Maier,~S.~A.; Giannini,~V.
  Quantum {Plasmonics}. \emph{Proceedings of the IEEE} \textbf{2016},
  \emph{104}, 2307--2322, Conference Name: Proceedings of the IEEE\relax
\mciteBstWouldAddEndPuncttrue
\mciteSetBstMidEndSepPunct{\mcitedefaultmidpunct}
{\mcitedefaultendpunct}{\mcitedefaultseppunct}\relax
\EndOfBibitem
\bibitem[Dubi and Sivan(2019)Dubi, and Sivan]{dubiHotElectronsMetallic2019}
Dubi,~Y.; Sivan,~Y. ``{{Hot}}'' Electrons in Metallic Nanostructures\textemdash
  Non-Thermal Carriers or Heating? \emph{Light: Science \& Applications}
  \textbf{2019}, \emph{8}, 89\relax
\mciteBstWouldAddEndPuncttrue
\mciteSetBstMidEndSepPunct{\mcitedefaultmidpunct}
{\mcitedefaultendpunct}{\mcitedefaultseppunct}\relax
\EndOfBibitem
\bibitem[Govorov and Zhang(2015)Govorov, and
  Zhang]{govorovKineticDensityFunctional2015}
Govorov,~A.~O.; Zhang,~H. Kinetic {{Density Functional Theory}} for {{Plasmonic
  Nanostructures}}: {{Breaking}} of the {{Plasmon Peak}} in the {{Quantum
  Regime}} and {{Generation}} of {{Hot Electrons}}. \emph{The Journal of
  Physical Chemistry C} \textbf{2015}, \emph{119}, 6181--6194\relax
\mciteBstWouldAddEndPuncttrue
\mciteSetBstMidEndSepPunct{\mcitedefaultmidpunct}
{\mcitedefaultendpunct}{\mcitedefaultseppunct}\relax
\EndOfBibitem
\bibitem[Govorov \latin{et~al.}(2013)Govorov, Zhang, and
  Gun'ko]{govorovTheoryPhotoinjectionHot2013}
Govorov,~A.~O.; Zhang,~H.; Gun'ko,~Y.~K. Theory of {{Photoinjection}} of {{Hot
  Plasmonic Carriers}} from {{Metal Nanostructures}} into {{Semiconductors}}
  and {{Surface Molecules}}. \emph{The Journal of Physical Chemistry C}
  \textbf{2013}, \emph{117}, 16616--16631\relax
\mciteBstWouldAddEndPuncttrue
\mciteSetBstMidEndSepPunct{\mcitedefaultmidpunct}
{\mcitedefaultendpunct}{\mcitedefaultseppunct}\relax
\EndOfBibitem
\bibitem[Saavedra \latin{et~al.}(2016)Saavedra, {Asenjo-Garcia}, and
  {Garc{\'i}a de Abajo}]{saavedraHotElectronDynamicsThermalization2016}
Saavedra,~J. R.~M.; {Asenjo-Garcia},~A.; {Garc{\'i}a de Abajo},~F.~J.
  Hot-{{Electron Dynamics}} and {{Thermalization}} in {{Small Metallic
  Nanoparticles}}. \emph{ACS Photonics} \textbf{2016}, \emph{3},
  1637--1646\relax
\mciteBstWouldAddEndPuncttrue
\mciteSetBstMidEndSepPunct{\mcitedefaultmidpunct}
{\mcitedefaultendpunct}{\mcitedefaultseppunct}\relax
\EndOfBibitem
\bibitem[Liu \latin{et~al.}(2018)Liu, Zhang, Link, and
  Nordlander]{liu_relaxation_2018}
Liu,~J.~G.; Zhang,~H.; Link,~S.; Nordlander,~P. Relaxation of
  {Plasmon}-{Induced} {Hot} {Carriers}. \emph{ACS Photonics} \textbf{2018},
  \emph{5}, 2584--2595\relax
\mciteBstWouldAddEndPuncttrue
\mciteSetBstMidEndSepPunct{\mcitedefaultmidpunct}
{\mcitedefaultendpunct}{\mcitedefaultseppunct}\relax
\EndOfBibitem
\bibitem[Rom{\'a}n~Castellanos \latin{et~al.}(2020)Rom{\'a}n~Castellanos, Kahk,
  Hess, and Lischner]{romancastellanosGenerationPlasmonicHot2020}
Rom{\'a}n~Castellanos,~L.; Kahk,~J.~M.; Hess,~O.; Lischner,~J. Generation of
  Plasmonic Hot Carriers from D-Bands in Metallic Nanoparticles. \emph{The
  Journal of Chemical Physics} \textbf{2020}, \emph{152}, 104111\relax
\mciteBstWouldAddEndPuncttrue
\mciteSetBstMidEndSepPunct{\mcitedefaultmidpunct}
{\mcitedefaultendpunct}{\mcitedefaultseppunct}\relax
\EndOfBibitem
\bibitem[Rossi \latin{et~al.}(2020)Rossi, Erhart, and
  Kuisma]{rossiHotCarrierGenerationPlasmonic2020a}
Rossi,~T.~P.; Erhart,~P.; Kuisma,~M. Hot-{{Carrier Generation}} in {{Plasmonic
  Nanoparticles}}: {{The Importance}} of {{Atomic Structure}}. \emph{ACS Nano}
  \textbf{2020}, \emph{14}, 9963--9971\relax
\mciteBstWouldAddEndPuncttrue
\mciteSetBstMidEndSepPunct{\mcitedefaultmidpunct}
{\mcitedefaultendpunct}{\mcitedefaultseppunct}\relax
\EndOfBibitem
\bibitem[Jin \latin{et~al.}(2023)Jin, Herran, Cortes, and
  Lischner]{jin_theory_2023}
Jin,~H.; Herran,~M.; Cortes,~E.; Lischner,~J. Theory of hot-carrier generation
  in bimetallic plasmonic catalysts. 2023;
  \url{http://arxiv.org/abs/2306.02477}, arXiv:2306.02477 [cond-mat]\relax
\mciteBstWouldAddEndPuncttrue
\mciteSetBstMidEndSepPunct{\mcitedefaultmidpunct}
{\mcitedefaultendpunct}{\mcitedefaultseppunct}\relax
\EndOfBibitem
\bibitem[Al-Zubeidi \latin{et~al.}(2023)Al-Zubeidi, Wang, Lin, Flatebo, Landes,
  Ren, and Link]{al-zubeidi_d-band_2023}
Al-Zubeidi,~A.; Wang,~Y.; Lin,~J.; Flatebo,~C.; Landes,~C.~F.; Ren,~H.;
  Link,~S. d-{Band} {Holes} {React} at the {Tips} of {Gold} {Nanorods}.
  \emph{The Journal of Physical Chemistry Letters} \textbf{2023}, \emph{14},
  5297--5304\relax
\mciteBstWouldAddEndPuncttrue
\mciteSetBstMidEndSepPunct{\mcitedefaultmidpunct}
{\mcitedefaultendpunct}{\mcitedefaultseppunct}\relax
\EndOfBibitem
\bibitem[Bykov \latin{et~al.}(2021)Bykov, Roth, Sartorello, Salmón-Gamboa, and
  Zayats]{bykov_dynamics_2021}
Bykov,~A.~Y.; Roth,~D.~J.; Sartorello,~G.; Salmón-Gamboa,~J.~U.; Zayats,~A.~V.
  Dynamics of hot carriers in plasmonic heterostructures. \emph{Nanophotonics}
  \textbf{2021}, \emph{10}, 2929--2938, Publisher: De Gruyter\relax
\mciteBstWouldAddEndPuncttrue
\mciteSetBstMidEndSepPunct{\mcitedefaultmidpunct}
{\mcitedefaultendpunct}{\mcitedefaultseppunct}\relax
\EndOfBibitem
\bibitem[Brown \latin{et~al.}(2016)Brown, Sundararaman, Narang, Goddard, and
  Atwater]{brown_nonradiative_2016}
Brown,~A.~M.; Sundararaman,~R.; Narang,~P.; Goddard,~W. A.~I.; Atwater,~H.~A.
  Nonradiative {Plasmon} {Decay} and {Hot} {Carrier} {Dynamics}: {Effects} of
  {Phonons}, {Surfaces}, and {Geometry}. \emph{ACS Nano} \textbf{2016},
  \emph{10}, 957--966, Publisher: American Chemical Society\relax
\mciteBstWouldAddEndPuncttrue
\mciteSetBstMidEndSepPunct{\mcitedefaultmidpunct}
{\mcitedefaultendpunct}{\mcitedefaultseppunct}\relax
\EndOfBibitem
\bibitem[Kiani \latin{et~al.}(2023)Kiani, Bowman, Sabzehparvar, Karaman,
  Sundararaman, and Tagliabue]{kiani_interfacial_2023}
Kiani,~F.; Bowman,~A.~R.; Sabzehparvar,~M.; Karaman,~C.~O.; Sundararaman,~R.;
  Tagliabue,~G. Interfacial {Hot} {Carrier} {Collection} {Controls} {Plasmonic}
  {Chemistry}. 2023; \url{http://arxiv.org/abs/2307.09324}, arXiv:2307.09324
  [cond-mat, physics:physics]\relax
\mciteBstWouldAddEndPuncttrue
\mciteSetBstMidEndSepPunct{\mcitedefaultmidpunct}
{\mcitedefaultendpunct}{\mcitedefaultseppunct}\relax
\EndOfBibitem
\bibitem[Ramakrishnan \latin{et~al.}(2021)Ramakrishnan, Mohammadparast, Dadgar,
  Mou, Le, Wang, Jain, and Andiappan]{ramakrishnan_photoinduced_2021}
Ramakrishnan,~S.~B.; Mohammadparast,~F.; Dadgar,~A.~P.; Mou,~T.; Le,~T.;
  Wang,~B.; Jain,~P.~K.; Andiappan,~M. Photoinduced {Electron} and {Energy}
  {Transfer} {Pathways} and {Photocatalytic} {Mechanisms} in {Hybrid}
  {Plasmonic} {Photocatalysis}. \emph{Advanced Optical Materials}
  \textbf{2021}, \emph{9}, 2101128\relax
\mciteBstWouldAddEndPuncttrue
\mciteSetBstMidEndSepPunct{\mcitedefaultmidpunct}
{\mcitedefaultendpunct}{\mcitedefaultseppunct}\relax
\EndOfBibitem
\bibitem[Wei{\ss}e \latin{et~al.}(2006)Wei{\ss}e, Wellein, Alvermann, and
  Fehske]{weisseKernelPolynomialMethod2006}
Wei{\ss}e,~A.; Wellein,~G.; Alvermann,~A.; Fehske,~H. The Kernel Polynomial
  Method. \emph{Reviews of Modern Physics} \textbf{2006}, \emph{78},
  275--306\relax
\mciteBstWouldAddEndPuncttrue
\mciteSetBstMidEndSepPunct{\mcitedefaultmidpunct}
{\mcitedefaultendpunct}{\mcitedefaultseppunct}\relax
\EndOfBibitem
\bibitem[Silver(1994)]{silverDensitiesStatesMegaDimensional1994}
Silver,~R.~N. Densities of {{States}} of {{Mega-Dimensional Hamiltonian
  Matrices}}. \emph{Int. J. Mod. Phys. C} \textbf{1994}, \emph{5}, 935\relax
\mciteBstWouldAddEndPuncttrue
\mciteSetBstMidEndSepPunct{\mcitedefaultmidpunct}
{\mcitedefaultendpunct}{\mcitedefaultseppunct}\relax
\EndOfBibitem
\bibitem[Jo{\~a}o \latin{et~al.}(2022)Jo{\~a}o, Lopes, and
  Ferreira]{joaoHighresolutionRealspaceEvaluation2022}
Jo{\~a}o,~S.~M.; Lopes,~J. M. V.~P.; Ferreira,~A. High-Resolution Real-Space
  Evaluation of the Self-Energy Operator of Disordered Lattices: {{Gade}}
  Singularity, Spin\textendash Orbit Effects and p-Wave Superconductivity.
  \emph{Journal of Physics: Materials} \textbf{2022}, \relax
\mciteBstWouldAddEndPunctfalse
\mciteSetBstMidEndSepPunct{\mcitedefaultmidpunct}
{}{\mcitedefaultseppunct}\relax
\EndOfBibitem
\bibitem[Pires \latin{et~al.}(2021)Pires, Amorim, Ferreira, Adagideli,
  Mucciolo, and Lopes]{piresBreakdownUniversalityThreedimensional2021}
Pires,~J. P.~S.; Amorim,~B.; Ferreira,~A.; Adagideli,~I.; Mucciolo,~E.~R.;
  Lopes,~J. M. V.~P. Breakdown of Universality in Three-Dimensional {{Dirac}}
  Semimetals with Random Impurities. \emph{Physical Review Research}
  \textbf{2021}, \emph{3}, 013183\relax
\mciteBstWouldAddEndPuncttrue
\mciteSetBstMidEndSepPunct{\mcitedefaultmidpunct}
{\mcitedefaultendpunct}{\mcitedefaultseppunct}\relax
\EndOfBibitem
\bibitem[Ferreira and Mucciolo(2015)Ferreira, and
  Mucciolo]{PhysRevLett.115.106601}
Ferreira,~A.; Mucciolo,~E.~R. Critical {{Delocalization}} of {{Chiral Zero
  Energy Modes}} in {{Graphene}}. \emph{Phys. Rev. Lett.} \textbf{2015},
  \emph{115}, 106601\relax
\mciteBstWouldAddEndPuncttrue
\mciteSetBstMidEndSepPunct{\mcitedefaultmidpunct}
{\mcitedefaultendpunct}{\mcitedefaultseppunct}\relax
\EndOfBibitem
\bibitem[Jo{\~a}o \latin{et~al.}(2020)Jo{\~a}o, An{\dj}elkovi{\'c}, Covaci,
  Rappoport, Lopes, and Ferreira]{joaoKITEHighperformanceAccurate2020}
Jo{\~a}o,~S.~M.; An{\dj}elkovi{\'c},~M.; Covaci,~L.; Rappoport,~T.~G.;
  Lopes,~J. M. V.~P.; Ferreira,~A. {{KITE}}: High-Performance Accurate
  Modelling of Electronic Structure and Response Functions of Large Molecules,
  Disordered Crystals and Heterostructures. \emph{Royal Society Open Science}
  \textbf{2020}, \emph{7}, 191809\relax
\mciteBstWouldAddEndPuncttrue
\mciteSetBstMidEndSepPunct{\mcitedefaultmidpunct}
{\mcitedefaultendpunct}{\mcitedefaultseppunct}\relax
\EndOfBibitem
\bibitem[Weisse(2004)]{weisse2004chebyshev}
Weisse,~A. Chebyshev Expansion Approach to the {{AC}} Conductivity of the
  {{Anderson}} Model. \emph{The European Physical Journal B-Condensed Matter
  and Complex Systems} \textbf{2004}, \emph{40}, 125--128\relax
\mciteBstWouldAddEndPuncttrue
\mciteSetBstMidEndSepPunct{\mcitedefaultmidpunct}
{\mcitedefaultendpunct}{\mcitedefaultseppunct}\relax
\EndOfBibitem
\bibitem[Santos~Pires \latin{et~al.}(2022)Santos~Pires, Jo{\~a}o, Ferreira,
  Amorim, and Viana Parente~Lopes]{santospiresAnomalousTransportSignatures2022}
Santos~Pires,~J.~P.; Jo{\~a}o,~S.~M.; Ferreira,~A.; Amorim,~B.; Viana
  Parente~Lopes,~J.~M. Anomalous {{Transport Signatures}} in {{Weyl
  Semimetals}} with {{Point Defects}}. \emph{Physical Review Letters}
  \textbf{2022}, \emph{129}, 196601\relax
\mciteBstWouldAddEndPuncttrue
\mciteSetBstMidEndSepPunct{\mcitedefaultmidpunct}
{\mcitedefaultendpunct}{\mcitedefaultseppunct}\relax
\EndOfBibitem
\bibitem[Jo{\~a}o and Lopes(2019)Jo{\~a}o, and
  Lopes]{joaoBasisindependentSpectralMethods2019}
Jo{\~a}o,~S.~M.; Lopes,~J. M. V.~P. Basis-Independent Spectral Methods for
  Non-Linear Optical Response in Arbitrary Tight-Binding Models. \emph{Journal
  of Physics: Condensed Matter} \textbf{2019}, \emph{32}, 125901\relax
\mciteBstWouldAddEndPuncttrue
\mciteSetBstMidEndSepPunct{\mcitedefaultmidpunct}
{\mcitedefaultendpunct}{\mcitedefaultseppunct}\relax
\EndOfBibitem
\bibitem[Covaci \latin{et~al.}(2010)Covaci, Peeters, and
  Berciu]{covaciEfficientNumericalApproach2010}
Covaci,~L.; Peeters,~F.~M.; Berciu,~M. Efficient {{Numerical Approach}} to
  {{Inhomogeneous Superconductivity}}: {{The Chebyshev-Bogoliubov}}\textendash
  de {{Gennes Method}}. \emph{Physical Review Letters} \textbf{2010},
  \emph{105}, 167006\relax
\mciteBstWouldAddEndPuncttrue
\mciteSetBstMidEndSepPunct{\mcitedefaultmidpunct}
{\mcitedefaultendpunct}{\mcitedefaultseppunct}\relax
\EndOfBibitem
\bibitem[{An{\dj}elkovi {\'c}} \latin{et~al.}(2018){An{\dj}elkovi {\'c}},
  Covaci, and Peeters]{PhysRevMaterials.2.034004}
{An{\dj}elkovi {\'c}},~M.; Covaci,~L.; Peeters,~F.~M. {{DC}} Conductivity of
  Twisted Bilayer Graphene: {{Angle-dependent}} Transport Properties and
  Effects of Disorder. \emph{Phys. Rev. Materials} \textbf{2018}, \emph{2},
  034004\relax
\mciteBstWouldAddEndPuncttrue
\mciteSetBstMidEndSepPunct{\mcitedefaultmidpunct}
{\mcitedefaultendpunct}{\mcitedefaultseppunct}\relax
\EndOfBibitem
\bibitem[Canonico \latin{et~al.}(2018)Canonico, Garc{\'i}a, Rappoport,
  Ferreira, and Muniz]{canonicoShubnikovHaasOscillations2018}
Canonico,~L.~M.; Garc{\'i}a,~J.~H.; Rappoport,~T.~G.; Ferreira,~A.;
  Muniz,~R.~B. Shubnikov--de {{Haas}} Oscillations in the Anomalous {{Hall}}
  Conductivity of {{Chern}} Insulators. \emph{Physical Review B} \textbf{2018},
  \emph{98}, 085409\relax
\mciteBstWouldAddEndPuncttrue
\mciteSetBstMidEndSepPunct{\mcitedefaultmidpunct}
{\mcitedefaultendpunct}{\mcitedefaultseppunct}\relax
\EndOfBibitem
\bibitem[Jin \latin{et~al.}(2022)Jin, Kahk, Papaconstantopoulos, Ferreira, and
  Lischner]{jinPlasmonInducedHotCarriers2022}
Jin,~H.; Kahk,~J.~M.; Papaconstantopoulos,~D.~A.; Ferreira,~A.; Lischner,~J.
  Plasmon-{{Induced Hot Carriers}} from {{Interband}} and {{Intraband
  Transitions}} in {{Large Noble Metal Nanoparticles}}. \emph{PRX Energy}
  \textbf{2022}, \emph{1}, 013006\relax
\mciteBstWouldAddEndPuncttrue
\mciteSetBstMidEndSepPunct{\mcitedefaultmidpunct}
{\mcitedefaultendpunct}{\mcitedefaultseppunct}\relax
\EndOfBibitem
\bibitem[Haynes(2014)]{haynesCRCHandbookChemistry2014}
Haynes,~W.~M. \emph{{{CRC Handbook}} of {{Chemistry}} and {{Physics}}}, 95th
  ed.; {Taylor \& Francis}, 2014\relax
\mciteBstWouldAddEndPuncttrue
\mciteSetBstMidEndSepPunct{\mcitedefaultmidpunct}
{\mcitedefaultendpunct}{\mcitedefaultseppunct}\relax
\EndOfBibitem
\bibitem[Papaconstantopoulos(2015)]{papaconstantopoulosHandbookBandStructure2015}
Papaconstantopoulos,~D.~A. \emph{Handbook of the {{Band Structure}} of
  {{Elemental Solids}}}, 2nd ed.; {Springer}, 2015\relax
\mciteBstWouldAddEndPuncttrue
\mciteSetBstMidEndSepPunct{\mcitedefaultmidpunct}
{\mcitedefaultendpunct}{\mcitedefaultseppunct}\relax
\EndOfBibitem
\bibitem[Johnson and Christy(1972)Johnson, and Christy]{johnson_optical_1972}
Johnson,~P.~B.; Christy,~R.~W. Optical {Constants} of the {Noble} {Metals}.
  \emph{Physical Review B} \textbf{1972}, \emph{6}, 4370--4379\relax
\mciteBstWouldAddEndPuncttrue
\mciteSetBstMidEndSepPunct{\mcitedefaultmidpunct}
{\mcitedefaultendpunct}{\mcitedefaultseppunct}\relax
\EndOfBibitem
\bibitem[Link and {El-Sayed}(1999)Link, and
  {El-Sayed}]{linkSpectralPropertiesRelaxation1999}
Link,~S.; {El-Sayed},~M.~A. Spectral {{Properties}} and {{Relaxation Dynamics}}
  of {{Surface Plasmon Electronic Oscillations}} in {{Gold}} and {{Silver
  Nanodots}} and {{Nanorods}}. \emph{The Journal of Physical Chemistry B}
  \textbf{1999}, \emph{103}, 8410--8426\relax
\mciteBstWouldAddEndPuncttrue
\mciteSetBstMidEndSepPunct{\mcitedefaultmidpunct}
{\mcitedefaultendpunct}{\mcitedefaultseppunct}\relax
\EndOfBibitem
\bibitem[Voisin \latin{et~al.}(2000)Voisin, Christofilos, Del~Fatti, Vallée,
  Prével, Cottancin, Lermé, Pellarin, and Broyer]{voisin_size-dependent_2000}
Voisin,~C.; Christofilos,~D.; Del~Fatti,~N.; Vallée,~F.; Prével,~B.;
  Cottancin,~E.; Lermé,~J.; Pellarin,~M.; Broyer,~M. Size-{Dependent}
  {Electron}-{Electron} {Interactions} in {Metal} {Nanoparticles}.
  \emph{Physical Review Letters} \textbf{2000}, \emph{85}, 2200--2203\relax
\mciteBstWouldAddEndPuncttrue
\mciteSetBstMidEndSepPunct{\mcitedefaultmidpunct}
{\mcitedefaultendpunct}{\mcitedefaultseppunct}\relax
\EndOfBibitem
\bibitem[Arbouet \latin{et~al.}(2003)Arbouet, Voisin, Christofilos, Langot,
  Fatti, Vallée, Lermé, Celep, Cottancin, Gaudry, Pellarin, Broyer, Maillard,
  Pileni, and Treguer]{arbouet_electron-phonon_2003}
Arbouet,~A.; Voisin,~C.; Christofilos,~D.; Langot,~P.; Fatti,~N.~D.;
  Vallée,~F.; Lermé,~J.; Celep,~G.; Cottancin,~E.; Gaudry,~M.; Pellarin,~M.;
  Broyer,~M.; Maillard,~M.; Pileni,~M.~P.; Treguer,~M. Electron-{Phonon}
  {Scattering} in {Metal} {Clusters}. \emph{Physical Review Letters}
  \textbf{2003}, \emph{90}, 177401\relax
\mciteBstWouldAddEndPuncttrue
\mciteSetBstMidEndSepPunct{\mcitedefaultmidpunct}
{\mcitedefaultendpunct}{\mcitedefaultseppunct}\relax
\EndOfBibitem
\bibitem[Movsesyan \latin{et~al.}(2022)Movsesyan, Santiago, Burger,
  {Correa-Duarte}, Besteiro, Wang, and
  Govorov]{movsesyanPlasmonicNanocrystalsComplex2022}
Movsesyan,~A.; Santiago,~E.~Y.; Burger,~S.; {Correa-Duarte},~M.~A.;
  Besteiro,~L.~V.; Wang,~Z.; Govorov,~A.~O. Plasmonic {{Nanocrystals}} with
  {{Complex Shapes}} for {{Photocatalysis}} and {{Growth}}: {{Contrasting
  Anisotropic Hot-Electron Generation}} with the {{Photothermal Effect}}.
  \emph{Advanced Optical Materials} \textbf{2022}, \emph{10}, 2102663\relax
\mciteBstWouldAddEndPuncttrue
\mciteSetBstMidEndSepPunct{\mcitedefaultmidpunct}
{\mcitedefaultendpunct}{\mcitedefaultseppunct}\relax
\EndOfBibitem
\end{mcitethebibliography}

\end{document}